\documentstyle[11pt,newpasp,twoside]{article}
\def\edcomment#1{\iffalse\marginpar{\raggedright\sl#1\/}\else\relax\fi}
\catcode`\@=11
\def\gsim{\ifmmode{\mathrel{\mathpalette\@versim>}}
    \else{$\mathrel{\mathpalette\@versim>}$}\fi}
\def\lsim{\ifmmode{\mathrel{\mathpalette\@versim<}}
    \else{$\mathrel{\mathpalette\@versim<}$}\fi}
\def\@versim#1#2{\lower 2.9truept \vbox{\baselineskip 0pt \lineskip 
    0.5truept \ialign{$\m@th#1\hfil##\hfil$\crcr#2\crcr\sim\crcr}}}
\catcode`\@=12
\def\square{$\sqcap\hskip - 2.7  mm \sqcup$}
\def\squadg{\hbox{\square $^{\circ}$}}
\reversemarginpar

\begin{document}
\title{Future Opportunities for Globular Cluster Astronomy at Ground Based 
Facilities}
 \author{Alvio Renzini}
\affil{European Southern Observatory, D-85748, Garching, Germany}
\begin{abstract}
In spite of great progress over the last $\sim 10$ years, especially
thanks to HST, a number of exciting open problem still puzzle
astronomers working on globular clusters in our own and other
galaxies. These problems range from determining more accurate ages to
assess whether massive black holes hide at the center of some
clusters, from identifying the physical origin of red giant winds to
assess whether there is an influence of cluster structure and
dynamics on the evolution of individual cluster stars, from demonstrating
whether or not some clusters possess a dark matter halo on their own to
eventually understand the formation of globular clusters in the
context of galaxy formation, and more.
In this review, I briefly sketch how ground based telescopes and their
instrumentation can help solving these problems through the present decade, 
2001-2010. A glimpse to the next decade is also given.

\end{abstract}

\section{Introduction}

This conference demonstrates the wide, enduring scientific interest of
globular clusters (GC) for a broad variety of astrophysical and
cosmological issues. This is indeed a very active field of
astronomical research, and as such it has a number of open problems,
currently under investigation. In this context, I have been asked by
the organizers to review the perspectives for such problems to be
solved (or at least effectively attacked) in the near future using
ground based facilities, with emphasis on 8-10m class telescopes. I
will do so leaving implicit that space borne facilities will widely
complement in several areas, suffice to mention here the enormous
progress in GC research that has been achieved using HST and its
instruments. I should also acknowledge that this review is going to be
somewhat biased towards the ESO Very Large Telescope (VLT), because it
is the ground based facility I am more familiar with. Finally,
emphasis will especially be on the telescopes and instruments that
will be available in the course of the present decade.

In the next Section a list of open problems in GC research is presented, while
in the following Section 3 some of the tools that may solve them are briefly
mentioned and commented.

\section{What Are the Problems?}

Globular clusters offer what is perhaps a unique variety of research
opportunities. They are at once prototypical N-body systems for dynamical
studies and prototypical {\it simple stellar populations} for both stellar
evolution and population synthesis models. They are astronomical objects 
for which most accurate ages can be obtained and they are tracers of the 
potential well of their parent/host galaxies as well as  of their early 
chemical evolution and mass assembly. Here are some of the hot topics
in current GC research.

\begin{list}{$\bullet$}{\setlength{\leftmargin 7mm}%
\setlength{\itemindent -1.5mm}\setlength{\itemsep 0mm}%
\setlength{\topsep 0mm}\setlength{\parskip 0mm}}
\item {\bf Globular Cluster Ages} remain one of of the key issues in
observational cosmology (as a proxy to the age of the universe) and
can provide important clues on the formation and early evolution of
the Milky Way galaxy and its satellites.  Improvements in the accuracy
of cluster age determinations depends on improving upon three main
ingredients, namely:
\begin{list}{$\star$}{\setlength{\leftmargin 5mm}%
\setlength{\itemindent -1.5mm}\setlength{\itemsep 0mm}%
\setlength{\topsep 0mm}\setlength{\parskip 0mm}}
\item {\sl Distances}: most progress is expected to come from space
      observations combined with multiobject spectroscopy from the ground
      (coupling proper motions and radial velocities in the so-called {\it
      quasi-geometric method}). For some 
      methods (e.g., main sequence fitting to local subdwarfs) next item will 
      also help. 
\item {\sl Composition}: [Fe/H] and [$\alpha$/H] are required to
      derive accurate ages from the main sequence turnoff. The
      abundance of s- and r-process elements and composition {\it
      anomalies} can be investigated using the same data.
\item {\sl Theoretical Models}: there is still room for improving and
      testing the stellar evolution {\it clock} used for age
      determinations, an activity for ground-based theorists which is
      not covered by this review.
\end{list}
\item {\bf Globular Cluster Dynamics} is another area of active
      observational and theoretical research. Progress in
      understanding the dynamical evolution of the clusters will
      be achieved by addressing issues such as:
\begin{list}{$\star$}{\setlength{\leftmargin 5mm}%
\setlength{\itemindent -1.5mm}\setlength{\itemsep 0mm}%
\setlength{\topsep 0mm}\setlength{\parskip 0mm}}
\item {\sl Binaries}: determining their frequency, radial
      distribution, etc., is of prime importance for understanding the
      dynamical evolution of the clusters.
\item {\sl Tidal Tails}: mapping the ongoing evaporation and stripping of 
      stars from the clusters. 
\item {\sl Central Black Hole}: do some cluster possess a central, massive BH?
      If so, did it form in a single event or from the merging of many
      stellar mass BHs?
\item {\sl Dark Matter}: are (some) clusters embedded in a ``dark matter'' 
      halo of their own? While certainly not primordial, some globular clusters
      may be the remnant of a bigger entity ...
\end{list}
\item {\bf Long Standing Puzzles} that still remain unresolved are quite
      numerous, for example:
\begin{list}{$\star$}{\setlength{\leftmargin 5mm}%
\setlength{\itemindent -1.5mm}\setlength{\itemsep 0mm}%
\setlength{\topsep 0mm}\setlength{\parskip 0mm}}
\item {\sl Mass Loss on RGB \& AGB}: color-magnitude diagrams provide
       compelling evidence that GC stars lose $\sim 0.2\, M_\odot$
       during the RGB and additional $\sim 0.1\, M_\odot$ during the
       AGB.  Star to star variations in the total amount of mass that is lost
       are also required to account for the HB morphology. Yet, almost
       nothing is known about the physical mechanism responsible for
       the wind mass loss in red giants, other than it cannot be
       radiation pressure (on either molecules or dust).
\item {\sl HB Blue Tails, Gaps, Jumps, and Rotation} still puzzle
       astronomers. We don't know for sure the origin of any of these
       phenomena, and to which extent they are related to each
       other. For a long time it has been suspected that the cluster
       structure (density, dynamical history, etc.) may play a role in
       establishing at least some of these features. There is a
       potential link between cluster dynamics and the evolution of
       individual cluster stars, quite worth to be explored further.
\item {\sl Composition Anomalies} affecting CNO and other light
       elements, whose origin is still debated (primordial? early
       accretion from AGB ejecta? deep mixing? in a combination thereof?).
\end{list}
\item {\bf Exotic Objects}: there is plenty
      of them in GCs, including:
\begin{list}{$\star$}{\setlength{\leftmargin 5mm}%
\setlength{\itemindent -1.5mm}\setlength{\itemsep 0mm}%
\setlength{\topsep 0mm}\setlength{\parskip 0mm}}
\item {\sl LMXBs} have been discovered in great numbers by X-ray
       satellites, and many still await for their optical
       identification and characterization.
\item {\sl Binary Pulsars} are especially frequent in GCs, and offer a
       unique opportunity to add information on the outcome of massive
       star supernova explosions, as well on the role of massive
       binaries on the dynamical evolution of the clusters.
\item {\sl Cataclysmic Variables} are also found (in quiescence) in
       relatively large numbers in deep HST images, and most of them
       await for further monitoring and spectroscopic study.
\end{list}
\item {\bf Globular Cluster Systems} in external galaxies is an area
       where progress has been extremely fast in recent years. Besides
       being interesting objects on their own, these GCs are used to
       map the potential well of the parent galaxies, and are even more
       interesting as fossil relics that may shed light on the
       formation and early evolution of their host galaxies. Ground
       based observations can greatly help to attack these problems,
       namely determining their
\begin{list}{$\star$}{\setlength{\leftmargin 5mm}%
\setlength{\itemindent -1.5mm}\setlength{\itemsep 0mm}%
\setlength{\topsep 0mm}\setlength{\parskip 0mm}}
\item {\sl Ages \& Metallicities} from colors and integrated spectra,
       also relative to the field population of the host, and their
\item {\sl Dynamics} within the galaxy potential well, probing the dark matter
       distribution at large distances, and reveling the degree of their
       orbital anisotropy and angular momentum.
\end{list}
\item {\bf Globular Cluster Formation} remains as perhaps the ultimate,
       still unsolved problem in GC research. 
\begin{list}{$\star$}{\setlength{\leftmargin 5mm}%
\setlength{\itemindent -1.5mm}\setlength{\itemsep 0mm}%
\setlength{\topsep 0mm}\setlength{\parskip 0mm}}
\item {\sl Why in the Very Early Universe} CGs formed so numerous? If
       this was prompted by galaxy merging, does this imply that most
       of the merging leading the assembly of galactic spheroids took
       place at high redshift, say $z\gsim 3$, with just sporadic such
       events later on?
\item {\sl Why the Metallicity Distribution} of the GCs is prominent at 
       low metallicities, contrary  to that of the stars in the host spheroid?
\item {\sl Why Bimodality} in the GC metallicity distribution is prominent
      in some galaxies and absent in others?
\item {\sl Why the Disks} of most spiral galaxies appear to be and
       have been sterile for GC formation? [Galactic GCs are clearly
       part of the spheroid, nothing to do with the disk, either thin
       of thick, with the metal rich ones belonging to the Galactic
       bulge.] What are the masses, ages, and metallicities of the few
       bright clusters occasionally found in disks?
\item {\sl Why Instead  LMC}, a flattened dwarf irregular, is so actively
       forming $\sim 10^5\,M_\odot$ GCs?
\item {\sl Why the average metallicity} of the GC populations is systematically
      lower than the average metallicity of the spheroids to which they belong?
\end{list}
\end{list}
\section{What Are the Tools?}
Having listed the problems, now comes a list of tools on
optical/infrared ground based telescopes that can help solving them.
\bigskip
\begin{list}{$\bullet$}{\setlength{\leftmargin 7mm}%
\setlength{\itemindent -1.5mm}\setlength{\itemsep 0mm}%
\setlength{\topsep 0mm}\setlength{\parskip 0mm}}
\item {\bf Optical/IR Wide Field Imagers} ($\sim 1/4-2 \squadg$ field
      of view) such as e.g., CFHT/Megacam, VST/$\Omega$Cam,
      UKIRT/WFCAM, VISTA. One can expect these facilities to be extensively
      used for:      
\begin{list}{$\star$}{\setlength{\leftmargin 5mm}%
\setlength{\itemindent -1.5mm}\setlength{\itemsep 0mm}%
\setlength{\topsep 0mm}\setlength{\parskip 0mm}}
\item {\sl Star Selection and Astrometry} for multiobject spectrographs 
      (see below).
\item {\sl Mapping Tidal Tails} in Galactic globulars.
\item {\sl GCs in External Galaxies}, their identification, astrometry
       and multicolor photometry.
\end{list}
\item {\bf Low- \& Medium-Resolution ($R=200-3000$) high multiplex
     (MPX) Spectrographs}, e.g., GEMINI/GMOS, Keck/DEIMOS,
     VLT/VIMOS, etc. will offer a great variety of opportunities for
     GC research, including:
\begin{list}{$\star$}{\setlength{\leftmargin 5mm}%
\setlength{\itemindent -1.5mm}\setlength{\itemsep 0mm}%
\setlength{\topsep 0mm}\setlength{\parskip 0mm}}
\item {\sl Star Membership} in GCs via radial velocity selection,
     especially useful for GCs projected over very crowded stellar
     fields, such as the GCs belonging to the Galactic bulge.
\item {\sl Composition Anomalies}, such as those of CH, CN, NH, and
      related anomalies, in both giants and dwarfs.
\item {\sl Line Indices} for GCs in other galaxies (e.g., the Lick
      indices $<\!{\rm Fe}\!>$, Mg$_2$, H$\beta$, etc.) will be of
      prime importance to better characterize GC families, compared to
      colors alone. Systematic differences or similarities among the
      various GC families and their correlations with the properties
      of the host galaxies could provide important clues on both
      galaxy and GC formation.
\item 
\end{list}
\end{list}
\medskip
For example, an instrument such as VIMOS at ($R\simeq 2000$, MPX=200) 
will allow to observe {\bf thousands} of objects/night down to mag $\sim 22$.
\begin{list}{$\bullet$}{\setlength{\leftmargin 7mm}%
\setlength{\itemindent -1.5mm}\setlength{\itemsep 0mm}%
\setlength{\topsep 0mm}\setlength{\parskip 0mm}}
\item {\bf High-Resolution, High Multiplex Spectrographs}
     ($R=5000-40,000$), e.g., the VLT FLAMES feeding UVES ($R=47,000$;
     MPX=8) \& GIRAFFE ($R=7000-20,000$; MPX=130) will begin a new era
     for stellar spectroscopy, in particular for GC stars. For
     example, such a facility will allow to obtain:
\begin{list}{$\star$}{\setlength{\leftmargin 5mm}%
\setlength{\itemindent -1.5mm}\setlength{\itemsep 0mm}%
\setlength{\topsep 0mm}\setlength{\parskip 0mm}}
\item {\sl Radial Velocities} ($\pm$ few km/s) for {\bf thousands} of
      mag $\lsim 20$ stars/night, that will allow extensive membership 
      assignment and census of binaries; again RVs with better accuracy 
      ($\lsim 1$ km/s) for {\bf thousands} of
      mag $\lsim 18$ stars/night, allowing distance determinations by the
      quasi-geometric method, in combination with proper motions
      with HST.
\item {\sl Chemical Composition} (from S/N$\sim 100$ spectra) for {\bf
      hundreds} of mag $\sim 18$ stars/night, allowing the abundance of
      many interesting elements to be determined, e.g., Li, C, N, O, Na, Mg, 
      Al, Si, Ca, Fe, Sr, Ba, Eu, and perhaps even Th, and U.
\item {\sl H$\alpha$ Emission} and narrow circumstellar absorption lines for 
      {\bf
      dozens} of RGB/AGB stars/night with $R\simeq 47,000$, thus enabling the 
      investigation
      of mass loss processes, and
\item {\sl Rotational Velocities} \& composition anomalies for {\bf dozens} of
      EHB stars/night.
\end{list}
\item
{\bf Integral Field (3D) Spectrographs}, such as e.g., VIMOS/IFU,
      GIRAFFE/ARGUS, SINFONI, CIRPASS, etc. will allow unprecedented
      spectroscopic studies in GC cores, including:
\begin{list}{$\star$}{\setlength{\leftmargin 5mm}%
\setlength{\itemindent -1.5mm}\setlength{\itemsep 0mm}%
\setlength{\topsep 0mm}\setlength{\parskip 0mm}}
\item {\sl Stellar Dynamics} in cluster cores, e.g., checking for massive 
      BHs, easily finding any very fast star that might be there,
\item {\sl Monitor Transients}, such as cataclismics, LMXBs, etc.,
\item {\sl Discover Interesting Objects}, e.g., very hard binaries, etc.
\end{list}
\item {\bf Adaptive Optics} (AO) assisted imaging and spectroscopy 
      (e.g., Keck/AO; GEMINI/ALTAIR; VLT/NACO; SUBARU/AO; VLT/SINFONI) may
      not help much to reach deeper limiting magnitudes in sparsely 
      populated fields,
      but certainly will add further opportunities in very crowded fields,
      such as GC cores and distant GCs. For example, AO will provide
\begin{list}{$\star$}{\setlength{\leftmargin 5mm}%
\setlength{\itemindent -1.5mm}\setlength{\itemsep 0mm}%
\setlength{\topsep 0mm}\setlength{\parskip 0mm}}
\item {\sl Deep near-IR imaging and spectroscopy} of stars within
        $\sim 20''$ from cluster center,
\item {\sl Deep near-IR imaging} of clusters in M31.
\end{list}
\item
{\bf High-resolution near-IR spectroscopy}, with e.g., CRIRES at the VLT, 
     $R=100,000$ (with AO), $\sim 22,000$ (without AO) will open up new opportunities
     for spectroscopic studies of GC stars, for example: 
\begin{list}{$\star$}{\setlength{\leftmargin 5mm}%
\setlength{\itemindent -1.5mm}\setlength{\itemsep 0mm}%
\setlength{\topsep 0mm}\setlength{\parskip 0mm}}
\item
{\sl Chemical Composition} from near-IR lines and bands, complementing
     optical spectroscopy.
\item {\sl  Obscured Clusters} in the Galactic bulge may become accessible to
     high resolution spectroscopy of cluster members. 
\end{list}

\item
{\bf 50-100m Telescopes} (e.g. OWL) 1 mas resolution, limiting magnitude
$\sim 35$, may well allow to:
\begin{list}{$\star$}{\setlength{\leftmargin 5mm}%
\setlength{\itemindent -1.5mm}\setlength{\itemsep 0mm}%
\setlength{\topsep 0mm}\setlength{\parskip 0mm}}
\item 
{\sl Map the Evolution of GC Systems} up to $z=5$ and beyond, and finally
\item 
{\sl See Globular Clusters in Formation}, all the way to $z=5$ and beyond!
\end{list}
\end{list}

\section{Conclusions}

Given the capabilities of the ground based 8--10m class telescopes and
of their instrumentation it is quite easy to anticipate where major
progress in GC research is going to happen in the {\bf present decade}
(2001-2010).

I believe that we are about to witness a quantum jump in multiobject
spectroscopy of GC stars, at low, intermediate, and high spectral
resolution. There will be the capability to observe thousands and
thousands of stars, and the impact on GC studies will perhaps be
similar to what were for imaging the CCDs in the '80s and HST/WFPC2 in
the '90s. Science will range from the production of {\it clean} CMDs
for radial velocity members to detailed multi-element chemical
composition studies, from stellar rotational velocities to velocity
dispersion at large distances from cluster center checking for the
presence of ``dark matter'', and more, being all this for huge samples
of stars. Next, is a jump of similar size in multiobject spectroscopy
of globular cluster systems, with hundreds of clusters in dozens of
galaxies being observed.  It is likely that this quantitative, orders
of magnitude expansion in the number of observed stars and clusters
will result in qualitative progress as well, with many of the open
problems of today being happily closed. Then, qualitatively new
science may well come from the 3D spectroscopy of cluster cores, with
a good chance to make surprising discoveries besides establishing
whether or not there is a massive BH lurking at the center of some
clusters.

However, in such optimistic scenario one caveat is in order.
Experience has shown that when a new facility is offered, with vastly
expanded capabilities over similar instruments of the previous
generation, then it is relatively easy to conceive a scientifically
exciting proposal, to get it approved by a telescope time allocation
committee, and then to get the data. Nonetheless, experience has also
shown that astronomers that were successful in collecting the data may
get stuck with them because of the inadequate hardware, and especially
software tools, at disposal for the reduction and scientific analysis
of the data. Given their enormous data rate, this is of particular
concern for the high-multiplex, multiobject spectrographs and the 3D
spectrographs. Without a dedicated effort to prepare the necessary
database environment and software tools many scientifically valuable
data may remain virtually unused for a very long period of time.

In the {\bf next decade} (2010-2020), ground based astronomy will be
dominated by ALMA and the next generation of the optical/IR extremely
large telescopes, CELT, GSMT, ELT, OWL, and whatever. With such large
optical/IR telescopes it will be finally possible to directly map the
evolution of globular clusters in galaxies all the way to see them in
formation, and eventually stick on the wall a poster with a million
pixel picture of a $z=5$ galaxy, with all her young globulars around.

\end{document}